# Navigating Design Science Research in mHealth Applications:
# A Guide to Best Practices

Avnish Singh Jat, Tor-Morten Grønli and George Ghinea

*Abstract*— **The rapid proliferation of mobile devices and advancements in wireless technologies have given rise to a new era of healthcare delivery through mobile health (mHealth) applications. Design Science Research (DSR) is a widely used research paradigm that aims to create and evaluate innovative artifacts to solve real-world problems. This paper presents a comprehensive framework for employing DSR in mHealth application projects to address healthcare challenges and improve patient outcomes. We discussed various DSR principles and methodologies, highlighting their applicability and importance in developing and evaluating mHealth applications. Furthermore, we present several case studies to exemplify the successful implementation of DSR in mHealth projects and provide practical recommendations for researchers and practitioners.**

## MANAGERIAL RELEVANCE STATEMENT

This study illuminates the integral role of Design Science Research (DSR) in optimizing the design, implementation, and evaluation of mHealth applications. For stakeholders in mHealth initiatives, embracing the DSR approach is not just recommended but essential. It ensures that applications are firmly rooted in addressing genuine healthcare challenges, leveraging cutting-edge, evidence-backed practices, and are exposed to rigorous appraisal. Management professionals should note the criticality of consistently involving stakeholders, tailoring solutions to user-specific requirements, and accentuating user-centric outcomes. Furthermore, an uncompromised commitment to data ethics, robust security measures, and informed user consent emerges as non-negotiable. The study also emphasizes forward-thinking in mHealth design, advocating for adaptability and scalability to meet evolving healthcare landscapes. In summary, the presented research paves the way for managers to oversee developing pioneering mHealth solutions that resonate with user needs, ensuring lasting impact and value.

## I. INTRODUCTION

Technology integration into healthcare has led to a rapid transformation in how healthcare services are delivered, monitored, and maintained. Mobile health (mHealth) applications have emerged as a promising area in healthcare, offering potential solutions to improve patient outcomes, reduce healthcare costs, and address various healthcare challenges [10]. We critically evaluated various methodological approaches to ascertain the most effective framework for our study. Among these, the Design Science Research (DSR) methodology emerged as the most suitable for our purposes, particularly in the development of innovative mHealth solutions. While methodologies like the Context-Intervention-Mechanism-Outcome (CIMO) logic are excellent for understanding the effectiveness of interventions in specific contexts by analyzing mechanisms and outcomes, thereby forming design principles that contribute to the body of knowledge, our study primarily utilizes Design Science Research (DSR). Given our focus on the iterative, problem-focused development of innovative mHealth technologies, we employ DSR as it integrates the design, development, analysis, and evaluation of artifacts, facilitating practical problem-solving and substantial theoretical contributions. [34].

### A. Background of mHealth applications

The ongoing digital revolution has fundamentally reshaped the landscape of various sectors worldwide, with healthcare being no exception. In the era of smartphones and the Internet of Things (IoT), a new frontier for healthcare delivery and management has emerged, known as mobile health (mHealth). The term mHealth refers to the medical and public health practice supported by mobile devices and wireless technologies. These advancements have opened up opportunities for more convenient, accessible, and personalized healthcare delivery, paving the way for an overall transformation of healthcare systems globally [16]. The advent of mHealth applications is a significant offshoot of this transformation, rapidly gaining traction due to its potential to revolutionize healthcare services. These applications leverage the ubiquitous nature of mobile devices and the power of wireless technology to provide a wide array of healthcare services ranging from diagnostics to therapeutics, preventive care, and health monitoring. They serve as digital health tools that can empower individuals to manage their health better, provide healthcare professionals with real-time patient data, and extend the reach of health services to remote and underserved areas [17].

Mobile devices such as smartphones, tablets, and wearables have become an integral part of people's lives. Their widespread adoption has been a critical factor in the growth and development of mHealth applications. These devices, equipped with various sensors and connectivity features, allow for seamless integration of health applications, making healthcare more accessible and efficient. Additionally, advancements in wireless communication technologies have enabled real-time, remote health monitoring and telemedicine, making healthcare more flexible and patient-centered [18]. The mHealth application landscape is varied, offering services like telemedicine, which enables virtual consultations and reduces physical visits, and remote monitoring, which helps healthcare providers track patient health in real time to prevent hospital readmissions and enable prompt intervention. Other applications focus on patient education, medication adherence, and chronic disease management, offering interactive and personalized solutions to improve patient outcomes [12].

### B. Design Science Research in Information Systems

Design Science Research (DSR) is a research methodology that seeks to address real-world challenges through the design and evaluation of innovative artifacts. It functions as a problem-solving paradigm, actively engaging in the creation of new and improved solutions to real-life issues. When applied within the field of Information Systems, DSR facilitates the development and implementation of new technologies, processes, and methods. Its goal is to bolster the efficiency and effectiveness of information systems within organizations [1][2]. DSR follows a systematic, iterative process that allows for continual refinement based on feedback. Initially, it identifies existing problems that need solutions. It then delves into designing those solutions and assessing their effectiveness. Upon evaluation, the solutions can be iteratively refined, resulting in improved versions that more precisely address the initial problem. This cyclical process enables an ongoing refinement of the artifact, ensuring that it continually improves and adapts to the context in which it operates [3].

### C. The significance of DSR in mHealth application projects

Design Science Research (DSR) enriches the body of knowledge in the mHealth sector by systematizing the creation and assessment of technological solutions aimed at healthcare challenges. In the realm of mHealth, DSR facilitates the introduction of novel applications, systems, or models through a systematic and creative process. These artifacts are rigorously designed with a focus on specific health-related needs and contribute significantly to the domain by providing new tools for healthcare delivery. The key strength of DSR lies in its strong focus on the design and evaluation of artifacts tailored to meet the specific needs and requirements of different stakeholders, including users, patients, and healthcare providers. This user-centric design approach aligns perfectly with the principles of mHealth, which strives to create applications that are user-friendly, accessible, and beneficial to the end-user [7]. Adopting a DSR approach allows researchers and practitioners in the mHealth domain to pinpoint and address the most pressing issues various stakeholders face. It enables the development of innovative, practical solutions that cater directly to these identified issues [8]. Once these solutions (or 'artifacts') are created, DSR doesn't stop there. It mandates a rigorous evaluation of the effectiveness of these solutions. This rigorous evaluation is vital, ensuring that the designed solutions truly resolve the issues at hand and contribute to improving healthcare delivery. The iterative nature of DSR offers another significant advantage. It ensures that mHealth applications are not static but dynamic and continually improving. The applications can be refined based on user feedback and the latest scientific knowledge, allowing them to adapt to changing user needs, evolving technology, and advances in medical research. This iterative process promotes the development of applications that stay relevant and effective over time [12][13]. Beyond artifact creation, DSR advances the methodologies used in the development and testing of mHealth solutions, setting new benchmarks for best practices. It also brings forward sophisticated evaluation techniques that measure the effectiveness, efficiency, and user engagement of mHealth technologies, offering a more comprehensive understanding of their practical value. Theoretical contributions are another significant addition, where DSR fosters the development of new theories concerning technology adoption and user interaction within the healthcare environment. This, in turn, informs the design of user-centric mHealth solutions. Moreover, DSR in mHealth yields actionable insights and guidelines for the implementation and broad-scale application of mobile health technologies. These guidelines help navigate the complexities of adapting solutions across diverse cultural and linguistic landscapes, ensuring that mHealth tools are accessible and relevant to a broad user base. Finally, DSR emphasizes the socio-technical impact assessment of mHealth solutions, shedding light on their influence on the dynamics of healthcare delivery, patient outcomes, and the overarching goal of health equity. By integrating DSR into mHealth research, the field gains innovative technological tools and a deeper scientific understanding and evidence-based insights into effective technology integration in healthcare practices. This contributes to a holistic approach to enhancing health outcomes through the informed use of technology [14][33].

## II. DESIGN SCIENCE RESEARCH PRINCIPLES

Design Science Research (DSR) is a systematic, iterative approach focused on solving real-world problems through the creation and rigorous evaluation of innovative artifacts. DSR principles provide a structured framework to guide researchers in the development, implementation, and refinement of these artifacts, ensuring they meet criteria of effectiveness, practical applicability, and scientific rigor. This section details the core principles of DSR, which include the relevance cycle, addressing the practical significance of the problem; the design cycle, which involves the iterative development and enhancement of the artifact; and the rigor cycle, ensuring that the research methodology adheres to stringent scientific standards. Additionally, the DSR process model is discussed, outlining the systematic approach to conducting research under this paradigm [2][6][28].

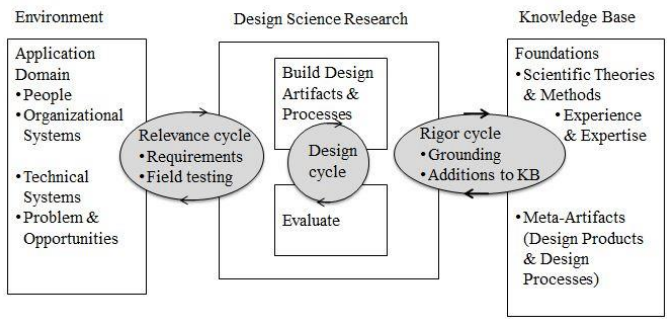

Figure 1. Design Science Research Cycles (A. R. Hevner, 2007)

## A. The Relevance Cycle

The relevance cycle ensures that the research problem and the proposed solution are grounded in a real-world context. This cycle's primary focus is on understanding the problem domain, recognizing the needs of stakeholders, and identifying the potential impact of the designed artifact.

The journey of the relevance cycle begins with Problem Identification. This initial stage demands researchers to immerse themselves in the problem, striving for a comprehensive understanding of the issues they aim to address. To achieve this, they engage in a multi-faceted immersion process involving observation of the current healthcare environment, direct engagement with diverse stakeholders, and a thorough review of existing literature. This approach not only helps in identifying the real-world challenges and opportunities within the mHealth domain but also ensures that the solutions developed are deeply rooted in actual user needs and practical healthcare scenarios [39]. The focus is on grasping the implications of the problem and appreciating its significance within the context of real-world applications. Once the problem has been identified and understood, the cycle proceeds to the stage of Requirements Analysis. During this phase, researchers delve into the needs and expectations of the stakeholders, who can range from users and patients to healthcare providers. This analysis is instrumental in shaping a robust understanding of the solution's requirements, with the solution needing to respond effectively to these defined needs.

The final step of the relevance cycle, Defining Objectives, follows logically from the identification of the problem and the analysis of requirements. At this point, researchers craft clear and specific objectives for the artifact that they intend to design. These objectives must be directly aligned with the identified problem and the requirements elicited from the stakeholders. It is through this careful alignment that the designed artifact can truly resonate with and respond to the needs of its real-world context.

## B. The Design Cycle

Following the Relevance Cycle in Design Science Research (DSR) comes the Design Cycle. This stage is the crucible where the creation, implementation, and evaluation of the artifact happen. It's a dynamic, iterative process that breathes life into the conceptual design and helps validate its real-world utility and impact. The design cycle begins with the crafting of the artifact's conceptual design. Researchers undertake this task by applying their understanding of the identified problem and stakeholders' requirements gathered from the Relevance Cycle. They create a blueprint that both addresses the problem and fulfills stakeholder needs, blending creativity with functionality in a practical design that can later be implemented. Once the conceptual design is ready, researchers then embark on the implementation of the artifact. They convert the blueprint into a tangible, functional entity. In the context of mHealth, the artifact might take the form of a mobile application developed using appropriate technologies and methodologies. This stage calls for practical skills, technological know-how, and a keen understanding of user interface design. The final stage of the design cycle is the evaluation of the artifact. It's here that the functional artifact undergoes rigorous testing to ascertain its effectiveness, usability, and impact. Evaluations can employ a variety of methods, such as usability testing, controlled experiments, and field trials, all depending on the nature of the artifact and the context of its use. Crucially, the feedback obtained from these evaluations then becomes the cornerstone for the refinement of the artifact in subsequent iterations, thereby closing the loop in the Design Cycle and preparing the ground for the next stage of DSR – the Rigor Cycle [2].

## C. The Rigor Cycle

The Rigor Cycle represents the final, crucial phase of Design Science Research (DSR). Its primary function is to ensure that the conducted research is rooted in existing scientific knowledge and ultimately contributes to the field's forward momentum. It offers a robust framework for researchers to incorporate insights from previous work, highlight the novelty of their artifacts, and communicate their research effectively. The Rigor Cycle commences with a comprehensive review of related work. Researchers plunge into the depth of existing literature, seeking out relevant theories, methods, and empirical findings that could inform both the design and evaluation of their artifact. This process helps ensure that the research is scientifically grounded, drawing from and adding to the rich tapestry of previous scholarly endeavors. Having deeply engaged with existing literature, researchers are then tasked with demonstrating the novelty of their artifact. It's here that they must show how their designed artifact either builds upon existing solutions or deviates from them in meaningful ways. In doing so, researchers not only affirm the relevance and practicality of their work but also its contribution to the expansion of knowledge in the field. Finally, The Rigor Cycle emphasizes the importance of clear and detailed reporting in research. Researchers are urged to document their processes, methods, findings, and the broader implications of their work. Such transparent reporting enables others in the field to replicate and build on existing studies, thereby contributing to the advancement of scientific knowledge. In Design Science Research (DSR), this approach is particularly valuable. DSR focuses on creating and evaluating innovative solutions, ranging from tangible technologies like software to conceptual frameworks and methodologies. This diverse output from DSR enriches scientific knowledge with design principles, development strategies, and evaluation techniques essential for addressing real-world challenges across various sectors [40].

Replication plays a pivotal role in DSR, acting as a vital tool for validation. It involves reenacting the development and assessment phases of an artifact to confirm the original results' reliability and applicability in different scenarios. This step not only reaffirms the initial findings but also broadens the artifact's relevance and adaptability. Figure 2 delineates a Design Science Research (DSR) replication project's evolution across three levels. Level I addresses a single research cycle with testing and redesign. Level II expands to two cycles, incorporating adaptations and updates from prior insights. Level III encompasses all three cycles, culminating in a recreation phase that synthesizes the knowledge and refinements from the previous iterative processes. Moreover, replication drives innovation within DSR by encouraging the modification and enhancement of existing artifacts to meet new challenges, thus fostering the generation of fresh insights. This iterative process is crucial for the evolution of design science, ensuring that the field continues to produce rigorously tested and scientifically sound contributions [41].

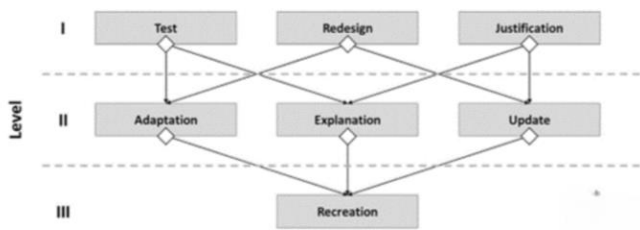

Figure 2. Progression framework for non-meta DSR replication study types (Alfred et al., 2021)

In domains such as mobile health (mHealth), the replication of DSR projects is crucial for refining solutions, identifying their limitations, and furthering technological and practical progress [42]. By adhering to the principles of the Rigor Cycle, researchers ensure their work serves as a foundation for future inquiries and innovations rather than existing in isolation. [2][35][36].

### D. The DSR process model

The Design Science Research (DSR) Process Model proposed by Vaishnavi et al. in 2019 seamlessly intertwines the relevance, design, and rigor cycles into a consolidated and comprehensive research approach. This integration showcases a broad view of the DSR methodology, emphasizing the dynamic and reciprocal relationships among the three cycles. These cycles are not merely stand-alone entities, but instead, they constantly interact and influence each other, making the overall process symbiotic and fluid. The DSR Process Model is distinctly iterative in nature. This implies that feedback obtained from the evaluation phase of the Design Cycle is actively employed to refine the produced artifact. This cycle of continuous refinement ensures that the artifact progressively evolves to better match the needs and requirements of the involved stakeholders. In the development of an artifact, requirements are pivotal, providing the foundational criteria that steer its creation. They delineate the artifact's intended functions, the problems it is designed to address, and the specific operational context it is meant to excel within. This groundwork is essential as it gives the design process its direction and purpose, ensuring that the resulting artifact is both relevant and targeted in its application. Once the artifact is realized, these same requirements become the benchmarks for evaluation, offering measurable standards to judge the artifact's performance and effectiveness. While needs and requirements are closely related, requirements being derived from needs are not the same. Needs are broader and more general, while requirements are detailed and specific. The distinction is crucial because it helps ensure that the artifact not only satisfies the stakeholders' immediate demands but does so in a way that is measurable and achievable, leading to a solution that effectively resolves the initial problem. Such adaptivity underscores the inherent flexibility of the DSR approach and its responsiveness to feedback and change [37].

When multiple artifacts are designed to address the same field problem using the same requirements in Design Science Research (DSR), the decision on which artifact to choose for detailed design and development often hinges on factors such as the artifact's potential impact, feasibility, innovativeness, and alignment with user needs. The decision-making process typically involves evaluating each artifact's suitability in terms of technical viability, resource availability, scalability, and the ability to meet the identified requirements effectively. Additionally, stakeholder feedback and empirical data gathered during the initial research phase play a crucial role in determining which solution offers the most practical and beneficial outcomes for the intended users. This approach ensures that the chosen artifact not only addresses the problem efficiently but also adds significant value to the field.

Figure 3 outlines the Design Science Research Methodology, which starts with the foundation of 'Knowledge Flows,' encompassing the theories, expertise, and information that permeate the research process. The 'Process Steps' begin with 'Problem Awareness,' where the research identifies and understands the problem at hand. Next is the 'Suggestion' phase, where a solution to the identified problem is hypothesized. This is followed by 'Development,' where the actual artifact, be it a model, method, or system, is created as a solution to the problem. Subsequently, the 'Evaluation' step involves assessing how effectively the artifact solves the problem. The final step is 'Conclusion,' where researchers draw conclusions based on the research process and artifact evaluation. The 'Deliverables' generated through this process include a 'Proposal' detailing the problem and proposed solution, a 'Tentative Design' of the artifact, the 'Artifact' itself, 'Performance Measures' to evaluate the artifact, and the 'Results,' which encapsulate the findings and performance evaluation of the artifact. [2][4][5].

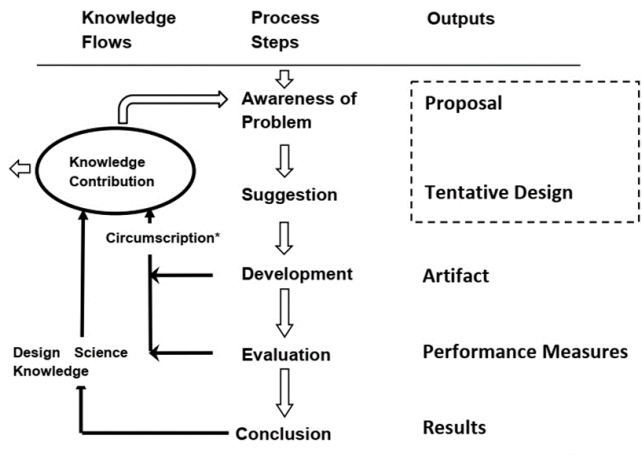

Fig. 3. The DSR process model (Vaishnavi et al., 2004)

An important aspect of this methodology is 'Circumscription,' a feedback loop indicating that researchers may revisit earlier stages to refine the solution if the artifact does not adequately solve the problem. Lastly, 'Operation and Goal Knowledge' is the process's bedrock, representing the research's practical understanding and objectives. This suggests a dynamic interaction where operational and goal-oriented knowledge guides the research and is enhanced by it. 'Design Science Knowledge' refers to the insights and understandings gained from building and assessing these artifacts. It contributes to the 'knowledge contribution' by providing novel solutions that are theoretically sound and empirically validated through their application in real-world scenarios. This knowledge is not just about the artifact itself but also about the design process, the methodologies applied, and the impact of the artifact within its environment. By traversing the stages from problem awareness to conclusion, DSR produces a rich understanding of both the problem space and the solution's effectiveness, thereby adding a valuable and often practical dimension to the existing body of knowledge within a field [27][38].

## III. METHODOLOGIES AND TECHNIQUES IN DSR FOR mHEALTH APPLICATIONS

Design Science Research (DSR) employs various methodologies and techniques to ensure the successful design, development, and evaluation of mHealth applications [9]. This section discusses the key methodologies and techniques used in DSR for mHealth applications, including requirements analysis, design and development approaches, and the importance of user-centered design, participatory design, and agile methodologies.

### A. Requirements Analysis

Requirements Analysis forms a pivotal step within the Design Science Research (DSR) framework. At this stage, researchers delve into the needs, expectations, and constraints of all stakeholders involved. In the context of mHealth applications, these stakeholders typically encompass users, patients, and healthcare providers. This in-depth analysis sets the stage for developing an artifact that effectively addresses real-world needs. In the data collection process for Design Science Research (DSR), researchers employ interviews, focus groups, surveys, and observations to thoroughly understand stakeholder needs and expectations [15]. During interviews and focus groups, questions are typically tailored to uncover stakeholders' experiences, challenges, preferences, and expectations from the proposed solution. These questions often stem from preliminary research or literature reviews. Surveys might include structured queries to quantify a larger population's needs, preferences, and patterns. Observational methods involve studying stakeholder interactions with current systems or environments, noting behaviors, bottlenecks, and usage patterns. This observational data is usually derived from theoretical frameworks or prior empirical studies. Together, these methods provide a comprehensive and multi-dimensional view of the requirements, informing the project's subsequent design and development phases. Once a substantial dataset is collected, researchers shift gears into data analysis. They scrutinize the amassed data to uncover patterns, trends, and relationships that help delineate the precise requirements of the mHealth application. This stage can often reveal nuanced insights and unspoken needs, which could prove instrumental in designing an artifact that truly resonates with its users. Following data analysis, researchers then prioritize the identified requirements. This step involves carefully considering factors such as the importance of each requirement, its feasibility within the project scope, and its impact on the project's overall objectives. In this way, Requirements Analysis ensures that the most critical stakeholder needs are brought to the forefront, setting the stage for the subsequent Design Cycle [1][15].

### B. Design and development

The design and development phase of DSR involves creating an artifact (e.g., a mHealth application) that addresses the identified problem and meets the stakeholders' requirements. This phase encompasses several design approaches, including user-centered design, participatory design, and agile methodologies [1][6].

#### 1) User-Centered Design

User-Centered Design (UCD) is an influential design philosophy that champions the inclusion of users at every step of the design journey. In the development of mHealth applications, UCD is a particularly effective approach, as it ensures that the final product is attuned to the unique needs, preferences, and expectations of its users.

A cornerstone of UCD is a profound understanding of the users. Researchers invest significant efforts in understanding users' characteristics, needs, and contexts. By exploring these facets in detail, researchers are better equipped to design a mHealth application that seamlessly aligns with users' expectations and daily lives. Researchers actively involve users in the design process once users' needs and contexts are comprehended. This inclusion is realized through various techniques, such as interviews, focus groups, and usability testing. This direct involvement allows for a first-hand insight into users' interactions with the product and their candid feedback. Such interaction uncovers hidden user needs and helps identify any pain points in the user experience.

UCD also encourages an iterative design approach. Rather than viewing design as a linear path leading to a final product, UCD perceives it as a cyclical process of continuous refinement. Researchers, guided by user feedback and evaluation results, iteratively tweak the design to enhance the mHealth application's effectiveness and usability. Through this series of refinements, UCD ensures that the final artifact isn't just an application but a comprehensive, user-friendly solution that substantially enhances the user's health management journey [17].

*2) Participatory Design*

Participatory Design is a synergistic design methodology that fosters a deep level of collaboration among users, healthcare providers, and other stakeholders throughout the design process. This hands-on involvement ensures that the resulting mHealth application is not just a mere technological creation but a relevant, practical, and effective tool for healthcare.

One of the significant techniques utilized in Participatory Design is the co-design workshop. These workshops present an engaging platform where researchers and stakeholders unite to share their diverse perspectives. Through a process of idea generation, prototype development, and solution exploration, these sessions foster a collaborative ethos that informs the design of the mHealth application.

Prototyping, another core element of Participatory Design, involves creating tangible representations of the mHealth application. Ranging from rudimentary, low-fidelity prototypes to more sophisticated, high-fidelity ones, these prototypes allow stakeholders to interact with the proposed design, providing valuable insights into its usability and practicality. Such feedback is then used to incrementally refine the design, aligning it more closely with stakeholder needs and contexts. Finally, Participatory Design emphasizes stakeholder validation. Researchers present the final design to the stakeholders, inviting them to review and validate it. This process ensures that the mHealth application reflects their needs, expectations, and contexts, reinforcing its practical value and effectiveness. Through its collaborative approach, Participatory Design brings the perspectives of all stakeholders into the design process, leading to mHealth applications that are not only technically sound but also deeply rooted in the practical realities of the users they are intended to serve [18].

*3) Agile Methodologies*

Agile methodologies represent a transformative approach to software development. They prioritize adaptability, teamwork, and continuous refinement, making them particularly suitable for mHealth application projects. The inherent flexibility of Agile methodologies allows researchers to adjust to shifting requirements, emerging technologies, and evolving contexts, ensuring that the final product remains responsive and relevant.

A key principle of Agile methodologies is incremental development. Instead of attempting to build the mHealth application as one massive undertaking, researchers break the task down into smaller, manageable parts. This incremental approach allows for ongoing enhancements and adaptations, responding nimbly to changes in requirements or emerging insights from user feedback. Another important principle is the emphasis on collaborative teamwork. Agile methodologies foster a close-knit working relationship among researchers, developers, users, and other stakeholders. This collaborative approach ensures that communication lines remain open, goals are aligned, and a sense of shared ownership permeates throughout the project [20].

Furthermore, Agile methodologies champion the value of regular feedback. Researchers make a concerted effort to gather feedback from users and stakeholders consistently. This feedback doesn't just guide the ongoing design and development of the mHealth application, but it also maintains an open dialogue with users and stakeholders, ensuring that their needs and experiences continue to inform the project's evolution. Through its commitment to adaptability, collaboration, and constant improvement, Agile methodologies offer a dynamic and responsive approach to mHealth application development. This ensures the end product remains attuned to its users' needs and the rapidly evolving healthcare technology landscape [19].

*C. Evaluation methods*

Evaluation is a critical aspect of the Design Science Research (DSR) process, as it helps researchers assess the effectiveness, usability, and impact of the designed mHealth application. Various evaluation methods can be employed to gather different types of data and insights, including usability testing, controlled experiments, field trials, and analytical evaluation techniques [6][21].

*1) Usability testing*

Usability testing is a method that consists of observing users while they navigate and interact with the mHealth application. This technique assists in pinpointing issues linked to the application's usability, its overall functionality, and the user's satisfaction level. Depending on the research aims, usability testing can be conducted in a controlled laboratory setting or a more organic, natural environment.

Within the scope of usability testing, task scenarios play an integral role. In these scenarios, researchers outline specific tasks that users are asked to carry out using the mHealth application, effectively simulating authentic use cases. This approach facilitates understanding how the application will likely be used in real-life situations. As users navigate the application, researchers closely observe their interactions, recording any difficulties, mistakes, or uncertainties encountered. This observational data provides crucial information about the application's user experience, helping identify potential improvement areas. After the testing session, researchers elicit feedback from the users. This process provides a deeper understanding of the users' experiences, their preferences, and any recommendations they might have for enhancing the application. Through this comprehensive approach, usability testing becomes a powerful tool in refining the mHealth application, ultimately leading to a more user-friendly and effective application to achieve its intended goals [22].

*2) Controlled experiments*

Controlled experiments are a key evaluation method wherein the mHealth application is compared to an alternative solution or control group. This comparison aims to assess various parameters such as its effectiveness, efficiency, or impact on specific outcomes. Typically, controlled experiments follow a rigorously defined protocol, which includes random assignment of participants to different conditions. Formulating a hypothesis is the first step in conducting a controlled experiment. Here, researchers establish expectations regarding the mHealth application's potential influence on specific outcomes. These outcomes could encompass a variety of factors, such as patient adherence to treatment or symptom management. The hypothesis serves as the guiding principle for the design and execution of the experiment. The next stage involves designing the experiment itself. This process includes selecting suitable participants, assigning them to specific conditions, and determining the measurement methods for the outcomes. Careful attention to detail in the design phase ensures the controlled experiment is well-structured and meaningful.

Following data collection, researchers embark on statistical analysis. This analysis aims to ascertain whether the mHealth application significantly influences the outcomes of interest relative to an alternative solution or control group. Findings from this analysis demonstrate the mHealth application's efficacy and provide valuable insights that can guide further development and refinement. Therefore, controlled experiments play an instrumental role in the DSR process, contributing to producing an effective and impactful mHealth application [23][24].

*3) Field trials*

Field trials constitute a vital component of the evaluation process, wherein the mHealth application is deployed in a real-world context. This allows for a practical assessment of its usability, effectiveness, and overall impact under naturally occurring conditions. By conducting field trials, researchers can gather valuable data on the application's performance in sync with the users' daily lives and routines, thus providing a more authentic evaluation. The first step in conducting field trials is deployment, where researchers release the mHealth application to a group of intended users or healthcare providers for a specified period. This gives them the opportunity to interact with the application in their day-to-day activities, mirroring real-world usage scenarios. Next is data collection, where researchers gather data on various aspects, such as user interaction with the application, their experiences, and the influence of the application on relevant outcomes. This data provides crucial insights into the practical functionality and effectiveness of the mHealth application, aiding in its further refinement. Finally, field trials offer researchers contextual insights. By observing the application in its intended setting, researchers understand how various contextual factors can influence its performance and impact. These factors can include user environments, social settings, routines, etc. Such understanding can guide researchers to make informed modifications to the mHealth application, optimizing its usability and effectiveness in real-world settings [24].

*4) Analytical evaluation techniques*

Analytical evaluation techniques involve systematically analyzing the mHealth application's design, functionality, and performance based on established models, frameworks, or principles. Examples of analytical evaluation techniques include heuristic evaluation, cognitive walkthroughs, and expert reviews. These techniques enable researchers to identify potential issues and areas for improvement without extensive user testing [25].

*D. Artifact Refinement and Iteration*

The artifact refinement and iteration process are crucial components of Design Science Research (DSR), as they pave the way for the continuous improvement of the mHealth application. This continuous improvement is guided by feedback received from evaluations and changing requirements that emerge during the project. The refinement and iteration process begins with incorporating feedback. Researchers utilize insights and feedback gathered from various evaluation methods to pinpoint areas within the mHealth application that require improvement. This feedback may relate to aspects such as usability, functionality, performance, or other specific features. Following this, the process of revising the design takes place. Based on the identified improvement areas, researchers make necessary changes to the application's design, functionality, or features. These revisions aim to address the issues raised in the feedback and enhance the mHealth application's overall performance. The goal is to create an application that aligns more closely with users' needs and expectations while providing effective solutions to identified health challenges. Lastly, the process involves iterative evaluation. Post-revision, researchers perform additional rounds of evaluation on the refined mHealth application. These evaluations aim to assess the effectiveness of the revisions and changes implemented. Moreover, they provide further feedback which fuels the ongoing process of refinement, thereby ensuring the application continuously evolves to meet the demands of its users and the dynamic landscape of mHealth. This iterative nature of DSR underscores its commitment to continual improvement, leading to the development of high-quality, impactful mHealth applications [26].

## IV. CASE STUDIES OF DSR IN MHEALTH APPLICATION PROJECTS

The following case studies demonstrate the application of Design Science Research (DSR) principles in the development and evaluation of mHealth applications. These examples highlight the value of DSR in creating effective and user-friendly mHealth solutions that address real-world healthcare challenges.

*A. Case Study 1: A telemedicine app for remote consultations for Patients with Sleep Apnea*

Researchers aimed to develop Ognomy, a telemedicine platform designed to facilitate consultations, diagnoses, and treatment for patients with sleep apnea. A Design Science Research (DSR) approach was used, initiating the process with a requirements analysis. To understand what was needed, researchers conducted structured interviews with six subject matter experts and held two brainstorming workshops. Through these interactive and thorough discussions, the following requirements emerged for the Ognomy telemedicine platform: patient registration and data collection, scheduling physician appointments, video consultation capabilities, and patient progress tracking. These requirements were translated into the design and development of four key artifacts: a mobile app for patients, a web app for providers, a reporting dashboard, and an AI-based chatbot for customer support and onboarding. The design process focused on creating a highly cohesive yet loosely coupled interaction among these components through a layered modular architecture using third-party APIs. Feedback from usability assessments led to further refinements, such as simplifying the onboarding process, enhancing status indicators during patient registration, and reorganizing the appointment calendar.

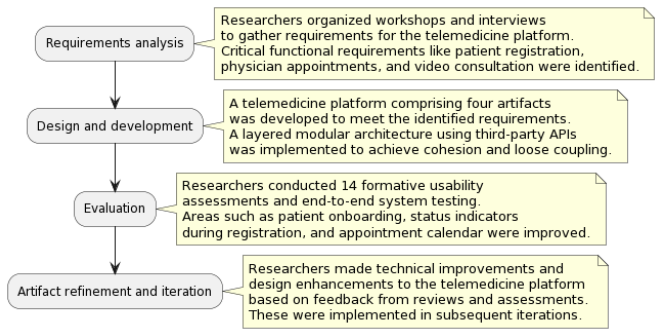

Fig.4 DSR Steps in Designing a Telemedicine App for Sleep Apnea Patients

Once the design phase was completed, the evaluation process commenced. Fourteen formative usability assessments were conducted, focusing on the patient onboarding process, status indicators during registration, and the appointment calendar layout. Further, end-to-end system testing was performed by three trained test engineers, providing a comprehensive evaluation of the entire platform. Lastly, refinement and iteration processes were undertaken. Researchers utilized feedback from both design reviews and usability assessments to make several key technical and design enhancements to the telemedicine platform. Implemented through subsequent iterations, these refinements ensured a more effective and user-friendly platform for sleep apnea patients and their healthcare providers, demonstrating the value and power of the DSR process in action [29].

## B. Case Study 2: A Health Coach-Augmented mHealth System for the Secondary Prevention of Coronary Heart Disease

Researchers aimed to create a home-based cardiac rehabilitation (HBCR) system. Their primary goal was to provide a robust solution for the self-management of chronic cardiovascular diseases and promote secondary prevention against other chronic illnesses sharing similar risk factors. To achieve this, they employed a Design Science Research (DSR) methodology, starting with identifying gaps in the current healthcare landscape. By conducting a comprehensive literature review and assessing various health facilities, they pinpointed gaps in current care practices for chronic conditions like hypertension and diabetes. It became evident that primary care settings often lacked evidence-based, integrated, and systematic management for these conditions, underscoring the need for a solution. To bridge the identified gaps, the researchers first delineated the requirements for the intervention, which were informed by consultations with IT experts, clinicians, and public health professionals. These requirements represented the essential functions and objectives the intervention needed to fulfill.

Based on these requirements, the researchers then developed the components of the HBCR intervention, functional elements such as integrated chronic condition management, clinical decision support grounded in evidence, comprehensive health data tracking, and an automated messaging system for enhancing medication adherence and appointment follow-ups. Subsequently, they developed the HBCR intervention, encompassing components such as integrated management of chronic conditions, evidence-based clinical decision support, longitudinal health data, and an automated short-messaging service to ensure compliance with medication regimens and follow-up visits. This resulted in a solution tailored to the specific needs of primary care settings, which could substantially improve chronic disease management. The HBCR intervention was then evaluated for its acceptability and feasibility through a pilot test involving ten coronary heart disease patients over 13 weeks.

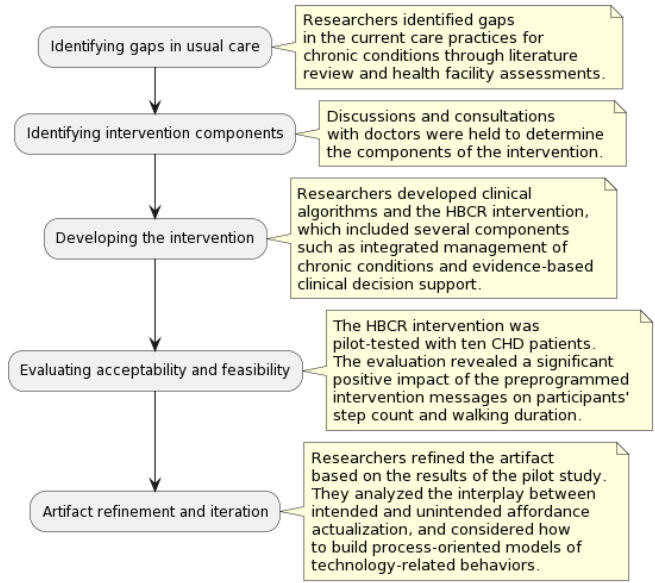

Fig.5 DSR Process for a Health Coach-Augmented mHealth System for Coronary Heart Disease Prevention

The pilot study revealed that preprogrammed intervention messages significantly boosted participants' daily step counts and walking duration, indicators positively associated with improved heart health. Taking the pilot study's results into account, the researchers refined and iterated the HBCR system. This involved analyzing the interaction between intended and unintended affordances, investigating unexpected actualizations, and building models of technology-related behaviors based on cognitive processes linked to technological affordance actualization. Through the successful application of the DSR framework, the study demonstrates the potential of the HBCR system to empower patients, provide tools and information for disease self-management, and highlight the importance of understanding both intended and unintended behaviors arising from technological interventions in chronic care settings.

The study demonstrates the successful application of a DSR framework in designing, developing, and evaluating a home-based cardiac rehabilitation system for chronic cardiovascular disease management. The system has the potential to empower patients with information, tools, and alerts and engage them in self-management of their diseases, while also considering the importance of unintended affordance actualization as a behavioral intervention for chronic care patients [30].

### C. Case Study 3: An mHealth App (Speech Banana) for Auditory Training

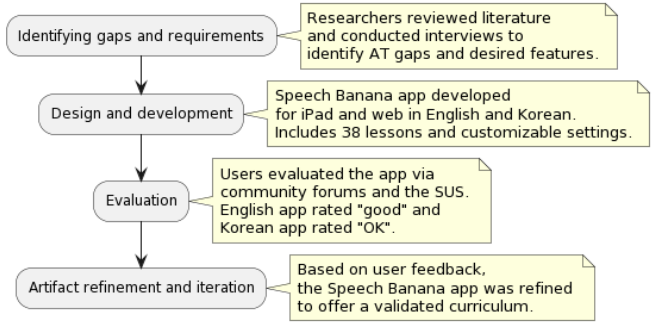

Fig.6 DSR Steps for the Speech Banana Auditory Training App Development

Researchers aimed to develop a mHealth application named Speech Banana. The application was intended for auditory training (AT) to serve the burgeoning demand among adults utilizing electronic hearing devices such as cochlear implants or hearing aids. To accomplish this, the team leveraged the Design Science Research Methodology (DSRM) approach. The initial stage involved identifying the existing gaps and requirements in AT. The researchers reviewed earlier literature and computer-based learning programs, outlining the current shortcomings in AT. They then interviewed speech pathologists and users to determine the app's necessary features.

After defining the requirements, the team proceeded to the design and development phase. They created the Speech Banana app for both iPad and web usage. Researchers have also prioritized both global accessibility and local relevance by developing the application in two languages: Korean and English. Korean was chosen as the local language in the region where the app was initially developed, ensuring it is accessible and user-friendly for a primary target audience. Meanwhile, they have also implemented an English version to extend the app's global reach and usability. The app consisted of 38 lessons, incorporating analytic exercises that paired visual and auditory stimuli and synthetic quizzes that exclusively presented auditory stimuli. It allowed users to modify settings such as speaker gender and background noise volume, enabling them to train with various frequencies and signal-to-noise ratios. To ascertain the app's effectiveness, the researchers invited former and current users to evaluate the application using community forums and the System Usability Scale (SUS). Six users rated the English version as "good", and sixteen users gave the Korean version an "OK" rating. Additionally, the English iPad app was downloaded over 3,200 times, and the Korean app was downloaded nearly 100 times, with over 100 users registering for the web apps. Based on these evaluations and user feedback, the Speech Banana app was refined to provide a validated curriculum that could help users develop speech comprehension skills via a mobile device. Thus, the study highlights the successful application of the DSRM approach in designing, developing, and evaluating a mHealth app for auditory training. The Speech Banana app could supplement clinical AT and enhance access to AT resources, especially in the global telemedicine context. The study demonstrates the successful application of a DSRM approach in designing, developing, and evaluating a mHealth app for auditory training. Speech Banana has the potential to supplement clinical AT and improve accessibility to AT resources, particularly in the context of global telemedicine [31].

### D. Case Study 4: Design Principles for mHealth Application Development in Rural Parts of Developing Countries: The Case of Noncommunicable Diseases in Kenya

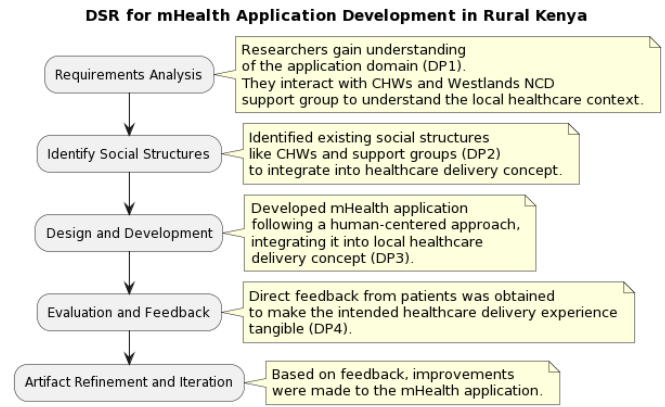

Fig.7 DSR Framework for mHealth App Design in Rural Kenya: Addressing Noncommunicable Diseases

This case study critically explores a Design Science Research (DSR) project based in rural Kenya that sought to address Non-Communicable Diseases (NCDs) via the development and implementation of a mobile health (mHealth) application. The project utilized the principles of a theory of design and action, which prescribes how to develop an artifact or devise a strategy for real-world application, as defined by Gregor in his typology of theories.

The study adhered to four crucial Design Principles (DPs), which guided the development and testing of multiple artifacts, all designed to aid rural communities in their struggle against NCDs. DP1 stresses the importance of developing a comprehensive understanding of the application domain before developing solutions. This principle proved pivotal during the project. The researchers' prior knowledge of Kenya's decentralized health system and early interactions with Community Health Workers (CHWs) and the Westlands NCD support group enabled them to create a solution that was not only accepted but also encouraged by the local communities. Understanding the country's specific healthcare challenges, including financial constraints and the essential role of CHWs, allowed researchers to avoid common pitfalls and focus on building a viable solution.

DP2 outlines the necessity of building upon existing social structures. The study identified two such structures: CHWs and support groups, which played significant roles in overcoming mobility and cost-related barriers within healthcare delivery. By embedding these structures within the healthcare delivery concept, the researchers could make a greater impact than they would have by only developing a mHealth application for individual patients.

DP3 underscores the critical importance of integrating the mHealth app into the local healthcare delivery concept. Adhering to a human-centered approach during the development of the mHealth app enabled the researchers to rapidly iterate and build an application that dovetailed seamlessly with existing processes. For instance, by accompanying CHWs on household visits, the research team could identify unique local challenges, such as network coverage issues, and ensure the application was compatible with offline usage.

DP4 focuses on rendering the intended healthcare delivery experience tangible for patients. Gaining direct feedback from patients was invaluable during the design process, informing a concept that was desirable and pragmatic for them. Developers were advised to keep in mind that patients often have a limited overview of the entirety of the healthcare delivery process, which can have implications for user testing.

While the case study demonstrated the successful application of DSR in the mHealth domain and its potential for improving healthcare delivery in resource-limited areas, it acknowledges certain limitations. The principles were only instantiated in three rural regions of Kenya and focused on NCDs, a set of diseases that affect a broad population. Future work will need to evaluate how well these principles can be generalized to other countries and specific target groups, and how they may be adjusted for other disease categories [32].

## V. PRACTICAL RECOMMENDATIONS FOR EMPLOYING DSR IN MHEALTH PROJECTS

Design Science Research (DSR) provides a systematic and iterative methodology that aids in creating user-friendly and effective mHealth applications. In discussing the application of Design Science Research (DSR) in mobile health (mHealth) development, the four case studies provide a rich tapestry of insights. DSR's contribution to the body of knowledge in mHealth is multifaceted, emphasizing user-centric design, contextual understanding, and integrating technology with healthcare practices. The development of the Ognomy app for sleep apnea, for instance, highlighted the importance of identifying specific user needs through direct engagement with stakeholders, creating a comprehensive telemedicine solution. Similarly, the Health Coach-Augmented mHealth System for

Coronary Heart Disease showcased DSR's ability to develop integrative solutions that combine various healthcare elements while also paying close attention to users' intended and unintended behaviors. The Speech Banana app for auditory training illustrated how DSR facilitates the creation of accessible, customizable, and widely applicable mHealth tools, emphasizing the significance of user feedback in the iterative design process. Lastly, the mHealth application for Non-Communicable Diseases in rural Kenya underscored the importance of adapting technology to local contexts, integrating it seamlessly with existing healthcare systems, and focusing on tangible patient experiences. Collectively, these case studies demonstrate how DSR contributes to creating innovative, practical, and sustainable mHealth solutions aligned with specific user needs and contexts, making a real impact in healthcare delivery. However, when utilizing DSR in mHealth projects, researchers must take into account a variety of practical elements in order to ensure successful outcomes. The subsequent recommendations provide a comprehensive guide to the primary considerations for undertaking DSR in mHealth projects:

### A. Forming Collaborative Relationships with Stakeholders

Establishing collaborative relationships with diverse stakeholders is fundamental when undertaking Design Science Research (DSR) for mHealth projects. Stakeholders can encompass various groups such as end-users, healthcare providers, and associated support organizations. Their active engagement is crucial to the process as it facilitates the design of a mHealth application that effectively addresses the real-world needs and expectations of its intended users in a comprehensive manner.

For fruitful collaboration to take place, it is vital that researchers actively involve stakeholders throughout every phase of the DSR process. Stakeholders' participation should not be confined to the final stages of design, development, and evaluation but should instead be integrated right from the outset of the process. In the early stages of problem identification, stakeholders, with their unique perspectives, can offer invaluable insights, contributing to a more accurate understanding of the issues to be addressed. Likewise, during the requirements analysis phase, their input can be instrumental in shaping the application's features and functions to align with the actual needs and preferences of the users.

As the project evolves, the stakeholders' role continues to be significant in providing valuable feedback and guidance, ensuring that the mHealth application remains true to its intended purpose and user needs. Their involvement in the design and development stages can contribute to creating an application that is user-friendly, pragmatic, and efficacious. Furthermore, during the evaluation stage, stakeholders can lend their expertise in assessing the application's effectiveness, usability, and overall impact, which could lead to further refinements and enhancements.

Maintaining successful stakeholder involvement throughout the DSR process necessitates the establishment of open and continuous communication channels. Regular interactions with stakeholders encourage a continuous exchange of ideas, feedback, and support, fostering a sense of mutual understanding among all parties involved. This ongoing engagement keeps the project on track and enables real-time modifications and improvements in response to feedback or changes in circumstances. Adopting this collaborative approach ensures that the mHealth application is technically competent, highly relevant, user-focused, and effective in addressing real-world needs. Consequently, initiating and preserving collaborative relationships with stakeholders emerge as a key aspect of successful DSR in mHealth projects.

### B. Striking a Balance between Innovation and Practicality

One of the central tenets of Design Science Research (DSR) is the pursuit of innovation. However, when conducting DSR in the context of mHealth projects, researchers must strike a careful balance between the desire for groundbreaking, innovative solutions and the practical constraints and requirements that these projects entail.

For instance, take the case study of the telemedicine platform Ognomy. In this project, the researchers used the DSR approach and started by building upon existing knowledge and established best practices within the telemedicine and sleep apnea treatment fields. By understanding the landscape thoroughly, including the technologies being employed, the successful interventions, and lessons from past projects, they developed a strong foundation for their work. This step was crucial as it helped avoid unnecessary repetition and potential obstacles, paving the way for the smooth progress of the project.

Once this solid foundational understanding was in place, the researchers began integrating innovative features into their solution. This wasn't innovation merely for the sake of novelty but a concerted effort to create groundbreaking solutions aligned with the stakeholders' needs and the project objectives. The researchers leveraged emerging technologies and employed novel design principles, as illustrated by creating four distinct artifacts (a mobile app for patients, a web app for providers, a dashboard for reporting, and an AI-based chatbot) within the Ognomy platform.

Yet, while they pursued innovation, the researchers never lost sight of practicality. They carefully considered the feasibility and sustainability of their proposed solution. Regarding feasibility, the researchers ensured their solution could be implemented given the available resources, technology, and operational environment. This consideration extended to the technical aspects like device compatibility, connectivity requirements, and software reliability, as well as logistical aspects such as project timelines and resource availability.

To navigate this delicate equilibrium, researchers should build upon existing knowledge and establish best practices within the mHealth domain. This approach involves a comprehensive understanding of the current landscape, including the technologies being used, successful interventions, and lessons learned from past projects. This foundation not only provides a solid base from which to work but also helps to avoid potential pitfalls and reinventing the wheel.

Once a strong foundational understanding is in place, researchers can begin to incorporate innovative features and functionalities into their mHealth solutions. The goal is not innovation for innovation's sake but rather the development of novel solutions that align with the needs of stakeholders and the project's objectives. This could involve leveraging emerging technologies, adopting novel design principles, or implementing cutting-edge data analytics capabilities, among other possibilities.

However, in the pursuit of innovation, practicality should not be overlooked. Feasibility and sustainability are two crucial aspects that researchers must consider when proposing solutions. Feasibility refers to the practicality of implementing the solution given the current resources, technology, and operational environment. This might involve considering technical aspects such as device compatibility, connectivity requirements, and software reliability, as well as logistical aspects such as project timelines and resource availability. Sustainability, on the other hand, pertains to the long-term viability of the solution. This involves considering factors such as cost (both initial and ongoing), user adoption, and maintenance requirements. For a mHealth solution to be truly sustainable, it must be economically viable, accepted by its users, and capable of being maintained and updated as necessary over time.

### C. Acknowledging and Addressing Ethical and Privacy Concerns

The use of mHealth applications often involves the collection, storage, and dissemination of sensitive health data, introducing a range of ethical and privacy concerns that researchers must attentively consider and address. Prioritizing the ethical management of user data and the preservation of privacy is not just a choice but a necessity that should be woven into the fabric of the mHealth project right from its design phase to implementation and evaluation. Firstly, researchers need to incorporate robust security and privacy protocols to protect users' sensitive health information. The security measures could encompass various techniques such as encryption to safeguard data during transmission and storage, setting up access controls to allow only authorized individuals to access the data, and establishing secure data storage systems to avoid potential data breaches. Furthermore, researchers may consider employing anonymization or de-identification techniques to further safeguard user identities by removing personal identifiers from the data. Beyond implementing technical security measures, the principle of informed consent is paramount. Researchers need to ensure that users are adequately informed about the purpose of data collection, the proposed use of their data, and the protective measures that are in place to preserve their privacy. Users should fully comprehend these aspects and voluntarily agree to them before their data is gathered. Informed consent respects user autonomy and empowers them with control over their personal information. Moreover, adhering to the relevant ethical guidelines, regulations, and standards that pertain to the use of health data is of utmost importance. These could include regulations like the Health Insurance Portability and Accountability Act (HIPAA) in the United States, the General Data Protection Regulation (GDPR) in the European Union, or any other regional or international regulations relevant to the project. These regulations establish a structured framework for data privacy and security, ensuring that mHealth applications uphold and protect user privacy to the highest degree possible. Ethical and privacy considerations are not optional facets of mHealth projects but rather fundamental requirements. By applying stringent security measures, obtaining informed consent, and adhering to relevant regulations, researchers can create mHealth applications that are not only effective and user-friendly but also respectful of ethical norms and protective of user privacy.

### D. Prioritizing Generalizability and Scalability

The fourth key consideration when undertaking DSR in mHealth projects is the generalizability and scalability of the mHealth application. To maximize the impact and reach of these applications, researchers should strive to develop solutions that can be applied to various populations, settings, and healthcare contexts. Furthermore, it is crucial to consider whether the application can scale effectively, accommodating user growth, functionalities, or geographic reach. To ensure generalizability, the mHealth application should be flexible and adaptable. This means creating an application customized to cater to different user groups, healthcare settings, or even different cultural contexts. A one-size-fits-all approach may not work in mHealth, as healthcare needs can greatly differ based on various factors such as geographical location, age group, or cultural background. Therefore, designing an application that can be easily adapted to these varying needs is crucial.

In addition to the design phase, researchers should also conduct evaluations using diverse participant samples and settings. This will enable them to assess the generalizability of the mHealth application and identify any potential adaptations that may be required for different contexts. Such evaluations can provide valuable insights into the real-world applicability and potential impact of the mHealth application across various demographic groups and healthcare settings. Scalability is another critical aspect that researchers must consider during the design and development phases. A mHealth application should be able to accommodate growth, whether in terms of user base, functionalities, or geographic reach. To plan for this, researchers should consider several factors, including the technological, organizational, and financial aspects of the application's growth. From a technological perspective, the infrastructure should be robust and capable of supporting an increased load as the application grows. Organizational considerations might include user support and training, ensuring that the application continues to offer a smooth and efficient user experience as the user base grows. Financially, funding models should be considered that can support the growth of the application, whether through investments, partnerships, or other revenue streams.

## VI. CONCLUSION

Design Science Research (DSR) offers a valuable framework for developing and evaluating mHealth applications, addressing real-world healthcare challenges, and meeting the needs of users and other stakeholders.[13] This paper has provided an overview of DSR principles, methodologies, techniques, and practical recommendations for conducting DSR in mHealth projects. The paper has outlined key principles of DSR, including the relevance cycle, the design cycle, and the rigor cycle. It also discusses various methodologies and techniques employed in DSR for mHealth applications, such as requirements analysis, design and development approaches, and evaluation methods. Furthermore, case studies of DSR in mHealth application projects have demonstrated the value of DSR in creating effective and user-friendly mHealth solutions. Finally, practical recommendations have been provided to guide researchers in employing DSR in mHealth projects.

This study highlights key points for those working in mobile health (mHealth). Firstly, it's essential to use the Design Science Research (DSR) method. This ensures mHealth apps meet real needs, are based on solid evidence, and get thoroughly tested. DSR encourages ongoing learning and innovation, leading to effective solutions. It's also important to involve all stakeholders in the mHealth project. This collaboration improves the app's relevance, ease of use, and effectiveness. Involving everyone, from users to healthcare providers, ensures the app meets real needs, leading to greater user satisfaction and success. Addressing ethical and privacy issues in mHealth apps is crucial. This means strong security, ethical data handling, and getting user consent. These steps build trust and confidence, which are key for app adoption and use. Lastly, the study stresses the need for mHealth apps to be adaptable and scalable. They should work for various groups and settings and be ready to grow and change with healthcare needs. This forward-thinking approach ensures the long-term relevance and sustainability of mHealth apps.

Future research in DSR for mHealth should delve into its applicability to cutting-edge healthcare technologies, such as artificial intelligence and blockchain [11]. Additionally, there's an opportunity to tackle global health challenges and combine DSR with other research paradigms, thus amplifying the impact of mHealth applications. Integrating CIMO logic (Context, Intervention, Mechanism, Outcome) in DSR methodologies is a pivotal aspect for future exploration. This integration could offer nuanced insights into the effectiveness of mHealth interventions in varied contexts, ensuring that the applications are innovative and highly effective in improving healthcare outcomes. By applying DSR principles and incorporating methodologies like CIMO logic, mHealth projects are poised to lead to groundbreaking applications with significant positive impacts on healthcare.

### ACKNOWLEDGMENT

I want to express my deepest gratitude to Kristiania University College for their unwavering support and invaluable resources, which have significantly contributed to the successful completion of this article.